\documentclass[twocolumn,twoside]{IEEEtran}
\usepackage{cite}      
\usepackage{subfigure} 
\usepackage{amsmath}   
\usepackage{array}
\usepackage{xfrac}

\usepackage{authblk}

\usepackage{balance}

\usepackage{graphicx}
\usepackage{amssymb}
\usepackage{tabularx}
\usepackage{amsmath}
\usepackage{booktabs}
\usepackage{graphicx}
\input{epsf}

\begin{document}

\title{Redundancy and Decoherence: Gateways From Quantum World to The Classical}

\author[ {\dag} ]{\rm Alireza Poostindouz}
\author[ {\dag} ]{\rm Vahid Salari}
\author[ {\ddag} ]{\rm Hamidreza Mohammadi}

\affil[{\dag} ]{{Department of Physics, Isfahan University of Technology, Isfahan, Iran}}
\affil[{\ddag} ]{{Department of Physics, Faculty of Science, University of Isfahan, Isfahan, Iran}}
\affil[ ]{\small{ poostindouz@ieee.org}}

\markboth{Third Conference on Recent Progress in Foundations of Physics, School of Physics, IPM, Tehran, 6-7 November 2013}{}
\pagenumbering{gobble}

\maketitle

\begin{abstract}
\/ In our  daily life experiences we face localized objects which are "here \emph{or} there" not "here \emph{and} there".  The state of a cat could be "dead and alive" at the same time from a quantum mechanical point of view, which is not in agreement with  our classical life observations. If we assume that quantum theory can explain the large scale events, how  is it possible that \emph{quantumness} disappears in the \emph{classical} world? The answer lies in the interaction between the environment and the system of interest. Quantum Darwinism describes the proliferation of the quantum information of each system in the environment. We will show how entanglement plays a key role in coding and decoding the information of each system in its surrounding environment.\\[1.5mm]
\end{abstract}

\IEEEpeerreviewmaketitle

\textbf{\textit{\small Keywords - Quantum Darwinism; Measurement Problem;\\ Redundancy; Decoherence; Quantum-to-Classical Transition}}

\section{Introduction}

Early decades of the twentieth century,  quantum theory was introduced and  its bizarre  effects were discovered in laboratories, which caused to change our attitude relative to the world around. The striking agreement between the experimental data and theoretical predictions of quantum theory made no choice for those who  did not believe the peculiar foundations, principles and consequences of quantum theory, to accept that the theory is the one which can describe the physical world at least at  microscopic level.  Albert Einstein tried his best to find a flaw in quantum mechanics, and his efforts  together with Podolsky and Rosen pointed out "\emph{A spooky action at distance}" which is nowadays exactly a core concept and basic notion in the quantum theory: Entanglement.

Entanglement, as a strong quantum correlation of two systems, is defined as losing of  the individuality of the subsystems,  which is why a valid physical description is a true explanation of the entangled systems as a whole. Based on quantum theory, this astounding effect reveals that  the local interactions may generate  distinctly nonlocal states. Thus, we may conclude that  there exist situations in quantum theory,  e.g. entangled systems, where the whole system is different from the sum of its subsystems.

As we will discuss here,  entanglement is a critical concept to explain decoherence which is due to an interaction with environment and is an answer to the measurement problem  (i.e., quantum-to-classical transition).

In the following section we will explain decoherence and will clarify how the entanglement between the environment and the system  causes a quantum coherent loss and the occurrence of super-selection phenomenon.  Then we will talk about the idea of Quantum Darwinism, which uses the role of redundancy and decoherence to explain the interaction between environment and the system of interest. Applications of redundancy can be found every where: like tracing down the history from the old ruins, having a reliable communication through wired and wireless networks, and the error free DNA transfer in organic processes.

%
%
\subsection{Quantum Mechanics Postulates}

 Here, we concentrate on four important axioms in quantum theory:
 \begin{enumerate}
   \item \emph{Superposition Principle}- The state of a system is represented by a vector in Hilbert space.
   \item Evolutions are unitary.
   \item Immediate repetition of a measurement results in the same outcome.
   \item \emph{Collapse Postulate}- Outcome of a measurement is only one of the eigenstates of the measured observable.
 \end{enumerate}

Based on the first postulate, as states are described in terms of vectors in a Hilbert space, a linear combination of quantum states is also a valid quantum state.

\begin{equation}
\left\vert\psi\right\rangle=\sum_n c_n\vert{\varphi_n}\rangle
\end{equation}

The collapse postulate which is inconsistent with the two first axioms is the core issue in the measurement problem.

\subsection{Entanglement And Bohr's Rule}

 Entanglement is described as a situation where we cannot write the state of the $\mathcal{SE}$ as a tensor product of the states of $\mathcal{S}$ and $\mathcal{E}$.

\begin{equation}
\left\vert\psi\right\rangle_{\mathcal{SE}}\neq \left\vert{\varphi_1}\right\rangle_\mathcal{S} \otimes \left\vert{\varphi_2}\right\rangle_{\mathcal{E}}
\end{equation}

One of the first examples of the entanglement was pointed out in the Schr\"{o}dinger's seminal paper, where the term "Entanglement" was firstly proposed for such quantum coupling of two systems.

According to the Bohr's rule, which was the first answer to the "quantum vs. classical paradox",  quantum theory is the best  formalism for microscopic world and the classical physics' laws are valid for macroscopic world.

The famous example, "Schr\"{o}dinger's Cat",  indicates that  it is possible to transfer quantum features to the classical realm via coupling a microscopic system to a macroscopic one.

\balance

\subsection{Measurement Problem}

In fact, the quantum measurement problem consists of three different problems; The Measurement Problems: First, the "Collapse Problem" which says  how an outcome is obtained out of all different possible outcomes? Second, the "Quantum to Classical Problem" which is, why quantum features disappears  at the macroscopic level? And finally, the "Preferred Basis Problem" which is why some physical quantities, like position, are mostly preferred rather than others?

Decoherence successfully answered the two last questions, while Quantum Darwinism tries to resolve the collapse paradox. \cite{Schlosshauer}

\section{Decoherence And Einselection}

As described before, if we assume that the two subsystems $\mathcal{A}$ and $\mathcal{B}$ are entangled we cannot deal with each system as an isolated system, thus there are no such vectors in the relevant Hilbert spaces to describe the states of individual parts. All the information we can obtain is the reduced density matrix which is:

\begin{equation}
{\hat{\rho}}_\mathcal{A}= \rm{Tr}_\mathcal{B}\left({\hat{\rho}}_{\mathcal{AB}}\right)
\end{equation}

The main idea that describes the "quantum to classical transition" is the environmental interactions with the system, which usually causes an entanglement in the "system-environment" state. The coherence of the system is brought to the entangled state as well, which makes the quantumness of the system inaccessible for the local observer.

Another consequence of the environment-system interaction is the super-selection phenomenon: The fact that some of the system states are sensitive to the environmental interactions and some other are more robust. These two consequences are the constituents of the concept of decoherence. \cite{Zurek2003}

\section{Error Correction Theory}

The error correction theory uses redundancy to make the messages immune from every type of noise. The application of this theory is ubiquitous even in the most natural domains. For instance, the key that makes it possible to trace and uncover the inscriptions of the remained historical ruins is the natural redundancy in almost all the human languages.

To explain the basic notion of the error correction, imagine the codes are $'0'$ and $'1'$. In order to send each bit you may just repeat it some time to make sure that it will correctly deliver to the receiver, i.e. $'00\cdots00'$ and $'11\cdots11'$. Thus you have mapped $2$ codes from a space of size $2$ to $2$ codewords from a space of the size $2^n$.

In other words, we can say that error correction uses the redundancy to map all possible parts of the data into a set of codewords that belong to a larger space, in a way that typical errors cannot change the message to that \emph{extent} which gets undecipherable. '\emph{Extent}' here refers to a topological distance measure in the larger space, which is usually a Cartesian distance, Hamming distance or sometimes Entropy distance. 

\section{Quantum Darwinism}

Quantum Darwinism describes that the environmental interactions copy the records of the system's state in the fractions of environment. Assume that the environment is consist of $N$ fractions: $\mathcal{E}_1,\mathcal{E}_2,\cdots,\mathcal{E}_N$. Here we must consider the correlations of the system with the fractions of the environment, thus we need the relevant reduced density matrix:

\begin{equation}
{\hat{\rho}}_\mathcal{SF}= \rm{Tr}_{\mathcal{E}/\mathcal{F}}\left({\hat{\rho}}_{\mathcal{SE}}\right)
\end{equation}

Where ${\mathcal{E}/\mathcal{F}}$ describes the entire environment except for the fraction $\mathcal{F}$.  The amount of information about $\mathcal{S}$ obtained by measuring the fraction $\mathcal{F}$ can be calculated by using mutual information:

\begin{equation}
I(\mathcal{S}:\mathcal{F}) = H_\mathcal{S} + H_\mathcal{F} - H_{\mathcal{S,F}}
\end{equation}

Based on the mutual information approach, we will notice that we can gain the required information bit by bit via measuring a few number of fractions of $\mathcal{E}$ in which the records of the state of $\mathcal{S}$ have been copied by environmental correlations. Therefore, we can probe the state of the system of interest without any prior knowledge of the $\mathcal{S}$.

In this sense, the new definition of redundancy would be the number of independent fractions of environment that supply almost all information about $\mathcal{S}$. \cite{Blume-Kohout2005}

\begin{equation}
R_\delta ={\Large\sfrac{1}{f_\delta}}
\end{equation}

This explains why photons are the most suitable environmental fractions which can deliver the total information of the system. Photons interact with almost all systems, while they do not interact with each other. On the contrary, since the air molecules scatter from one another, they cannot save the records of the state of the system properly and thus they don't make a good environmental fraction to supply the information we are looking for. \cite{Zurek2009}
%

\bibliographystyle{ieeetr}
\bibliography{QuantumDarwinism}

\end{document}